\documentclass[twocolumn,showpacs,preprintnumbers,amsmath,amssymb,prl,superscriptaddress]{revtex4}

\usepackage{graphicx} 
\usepackage{dcolumn}  
\usepackage{bm}       
\usepackage{epsfig}

\newcommand{\He}{^{4}\mbox{He}}

\begin{document}

\preprint{WM-05-109}

\title{Parity-Violating Electron Scattering from $^4$He and the 
Strange Electric Form Factor of the Nucleon}

%
%
\author{K.~A.~Aniol}
\affiliation{ \mbox{California State University, Los Angeles}, Los Angeles, California 90032, USA }

\author{D.~S.~Armstrong}
\affiliation{College of William and Mary, Williamsburg, Virginia 23187, USA}

\author{T.~Averett}
\affiliation{College of William and Mary, Williamsburg, Virginia 23187, USA}

\author{H.~Benaoum}
\affiliation{Syracuse University, Syracuse, New York 13244, USA} 

\author{P.~Y.~Bertin}
\affiliation{Universit\'{e} Blaise Pascal/CNRS-IN2P3, F-63177 Aubi\`ere, France }

\author{E.~Burtin} 
\affiliation{CEA Saclay, DAPNIA/SPhN, F-91191 Gif-sur-Yvette, France } 

\author{J.~Cahoon}
\affiliation{University of Massachusetts Amherst, Amherst, Massachusetts 01003, USA}

\author{G.~D.~Cates}
\affiliation{University of Virginia, Charlottesville, Virginia 22904, USA}
 
\author{C.~C.~Chang} 
\affiliation{University of Maryland, College Park, Maryland 20742, USA} 

\author{Y.-C.~Chao} 
\affiliation{Thomas Jefferson National Accelerator Facility, Newport News, Virginia 23606, USA} 

\author{J.-P.~Chen} 
\affiliation{Thomas Jefferson National Accelerator Facility, Newport News, Virginia 23606, USA} 

\author{Seonho~Choi}
\affiliation{Temple University, Philadelphia, Pennsylvania 19122, USA} 

\author{E.~Chudakov} 
\affiliation{Thomas Jefferson National Accelerator Facility, Newport News, Virginia 23606, USA} 

\author{B.~Craver} 
\affiliation{University of Virginia, Charlottesville, Virginia 22904, USA}

\author{F.~Cusanno}
\affiliation{INFN, Sezione Sanit\`a, 00161 Roma, Italy} 

\author{P.~Decowski}
\affiliation{Smith College, Northampton, Massachusetts 01063, USA}

\author{D.~Deepa} 
\affiliation{Old Dominion University, Norfolk, Virginia 23508, USA} 

\author{C.~Ferdi}
\affiliation{Universit\'{e} Blaise Pascal/CNRS-IN2P3, F-63177 Aubi\`ere, France }

\author{R.~J.~Feuerbach}
\affiliation{Thomas Jefferson National Accelerator Facility, Newport News, Virginia 23606, USA} 

\author{J.~M.~Finn} 
\affiliation{College of William and Mary, Williamsburg, Virginia 23187, USA}

\author{S.~Frullani}
\affiliation{INFN, Sezione Sanit\`a, 00161 Roma, Italy} 

\author{K.~Fuoti}
\affiliation{University of Massachusetts Amherst, Amherst, Massachusetts 01003, USA}

\author{F.~Garibaldi} 
\affiliation{INFN, Sezione Sanit\`a, 00161 Roma, Italy} 

\author{R.~Gilman} 
\affiliation{Rutgers, The State University of New Jersey, Piscataway, New Jersey 08855, USA} 
\affiliation{Thomas Jefferson National Accelerator Facility, Newport News, Virginia 23606, USA} 

\author{A.~Glamazdin} 
\affiliation{Kharkov Institute of Physics and Technology, Kharkov 310108, Ukraine} 

\author{V.~Gorbenko} 
\affiliation{Kharkov Institute of Physics and Technology, Kharkov 310108, Ukraine} 

\author{J.~M.~Grames}
\affiliation{Thomas Jefferson National Accelerator Facility, Newport News, Virginia 23606, USA} 

\author{J.~Hansknecht} 
\affiliation{Thomas Jefferson National Accelerator Facility, Newport News, Virginia 23606, USA} 

\author{D.~W.~Higinbotham} 
\affiliation{Thomas Jefferson National Accelerator Facility, Newport News, Virginia 23606, USA} 

\author{R.~Holmes} 
\affiliation{Syracuse University, Syracuse, New York 13244, USA} 

\author{T.~Holmstrom} 
\affiliation{College of William and Mary, Williamsburg, Virginia 23187, USA}

\author{T.~B.~Humensky}
\affiliation{University of Chicago, Chicago, Illinois 60637, USA}

\author{H.~Ibrahim}
\affiliation{Old Dominion University, Norfolk, Virginia 23508, USA} 

\author{C.~W.~de~Jager} 
\affiliation{Thomas Jefferson National Accelerator Facility, Newport News, Virginia 23606, USA} 

\author{X.~Jiang} 
\affiliation{Rutgers, The State University of New Jersey, Piscataway, New Jersey 08855, USA} 

\author{L.~J.~Kaufman}
\affiliation{University of Massachusetts Amherst, Amherst, Massachusetts 01003, USA}

\author{A.~Kelleher} 
\affiliation{College of William and Mary, Williamsburg, Virginia 23187, USA}

\author{A.~Kolarkar} 
\affiliation{University of Kentucky, Lexington, Kentucky 40506, USA}

\author{S.~Kowalski}
\affiliation{Massachusetts Institute of Technology, Cambridge, Massachusetts 02139, USA} 

\author{K.~S.~Kumar}
\affiliation{University of Massachusetts Amherst, Amherst, Massachusetts 01003, USA}
 
\author{D.~Lambert}
\affiliation{Smith College, Northampton, Massachusetts 01063, USA}

\author{P.~LaViolette}
\affiliation{University of Massachusetts Amherst, Amherst, Massachusetts 01003, USA}
 
\author{J.~LeRose} 
\affiliation{Thomas Jefferson National Accelerator Facility, Newport News, Virginia 23606, USA} 

\author{D.~Lhuillier} 
\affiliation{CEA Saclay, DAPNIA/SPhN, F-91191 Gif-sur-Yvette, France } 

\author{N.~Liyanage}
\affiliation{University of Virginia, Charlottesville, Virginia 22904, USA}
 
\author{D.~J.~Margaziotis} 
\affiliation{ \mbox{California State University, Los Angeles}, Los Angeles, California 90032, USA }

\author{M.~Mazouz} 
\affiliation{Laboratoire de Physique Subatomique et de Cosmologie,
38026 Grenoble, France}

\author{K.~McCormick} 
\affiliation{Rutgers, The State University of New Jersey, Piscataway, New Jersey 08855, USA} 

\author{D.~G.~Meekins} 
\affiliation{Thomas Jefferson National Accelerator Facility, Newport News, Virginia 23606, USA} 

\author{Z.-E.~Meziani} 
\affiliation{Temple University, Philadelphia, Pennsylvania 19122, USA} 

\author{R.~Michaels} 
\affiliation{Thomas Jefferson National Accelerator Facility, Newport News, Virginia 23606, USA} 

\author{B.~Moffit}
\affiliation{College of William and Mary, Williamsburg, Virginia 23187, USA}

\author{P.~Monaghan} 
\affiliation{Massachusetts Institute of Technology, Cambridge, Massachusetts 02139, USA} 

\author{C.~Munoz-Camacho}
\affiliation{CEA Saclay, DAPNIA/SPhN, F-91191 Gif-sur-Yvette, France } 

\author{S.~Nanda}
\affiliation{Thomas Jefferson National Accelerator Facility, Newport News, Virginia 23606, USA} 

\author{V.~Nelyubin}
\affiliation{University of Virginia, Charlottesville, Virginia 22904, USA}
\affiliation{St.Petersburg Nuclear Physics Institute of Russian Academy of Science, Gatchina, 188350, Russia}

\author{D.~Neyret} 
\affiliation{CEA Saclay, DAPNIA/SPhN, F-91191 Gif-sur-Yvette, France } 

\author{K.~D.~Paschke}
\affiliation{University of Massachusetts Amherst, Amherst, Massachusetts 01003, USA}

\author{M.~Poelker} 
\affiliation{Thomas Jefferson National Accelerator Facility, Newport News, Virginia 23606, USA} 

\author{R.~Pomatsalyuk} 
\affiliation{Kharkov Institute of Physics and Technology, Kharkov 310108, Ukraine} 

\author{Y.~Qiang}
\affiliation{Massachusetts Institute of Technology, Cambridge, Massachusetts 02139, USA} 

\author{B.~Reitz} 
\affiliation{Thomas Jefferson National Accelerator Facility, Newport News, Virginia 23606, USA} 

\author{J.~Roche} 
\affiliation{Thomas Jefferson National Accelerator Facility, Newport News, Virginia 23606, USA} 

\author{A.~Saha} 
\affiliation{Thomas Jefferson National Accelerator Facility, Newport News, Virginia 23606, USA} 

\author{J.~Singh}
\affiliation{University of Virginia, Charlottesville, Virginia 22904, USA}

\author{R.~Snyder}
\affiliation{University of Virginia, Charlottesville, Virginia 22904, USA}

\author{P.~A.~Souder}
\affiliation{Syracuse University, Syracuse, New York 13244, USA} 

\author{R.~Subedi}
\affiliation{Kent State University, Kent, Ohio 44242, USA} 

\author{R.~Suleiman} 
\affiliation{Massachusetts Institute of Technology, Cambridge, Massachusetts 02139, USA} 

\author{V.~Sulkosky}
\affiliation{College of William and Mary, Williamsburg, Virginia 23187, USA}

\author{W.~A.~Tobias}
\affiliation{University of Virginia, Charlottesville, Virginia 22904, USA}

\author{G.~M.~Urciuoli} 
\affiliation{INFN, Sezione Sanit\`a, 00161 Roma, Italy} 

\author{A.~Vacheret} 
\affiliation{CEA Saclay, DAPNIA/SPhN, F-91191 Gif-sur-Yvette, France } 

\author{E.~Voutier} 
\affiliation{Laboratoire de Physique Subatomique et de Cosmologie,
38026 Grenoble, France}

\author{K.~Wang} 
\affiliation{University of Virginia, Charlottesville, Virginia 22904, USA}

\author{R.~Wilson} 
\affiliation{Harvard University, Cambridge, Massachusetts 02138, USA} 

\author{B.~Wojtsekhowski}
\affiliation{Thomas Jefferson National Accelerator Facility, Newport News, Virginia 23606, USA} 
 
\author{X.~Zheng}
\affiliation{Argonne National Laboratory, Argonne, Illinois, 60439, USA}

\collaboration{The HAPPEX Collaboration}
\noaffiliation
%
%

\date{June 7, 2005} 

\begin{abstract}
We have measured the parity-violating
electroweak asymmetry in the elastic 
scattering of polarized electrons from $^4$He at an average 
scattering angle $\langle \theta_{lab} \rangle = 5.7^{\circ}$ and a 
four-momentum transfer $Q^2 = 0.091\; {\rm GeV}^2$. 
From these data, for the first time, the strange electric 
form factor of the nucleon $G^s_E$ can be isolated. 
The measured asymmetry of $A_{\rm PV} = (6.72 \pm 0.84_{\rm (stat)}
\pm 0.21_{\rm (syst)}) \times 10^{-6}$ yields 
a value of
$G^s_E = -0.038 \pm 0.042_{\rm (stat)} \pm  0.010_{\rm (syst)}
$, consistent with zero. 
\end{abstract}

\maketitle

The complex structure of the nucleon goes well beyond its
simplest description as a collection of three valence quarks.
The sea of gluons
and $q\overline{q}$ pairs that arise in quantum chromodynamics can
play an important role, 
possibly even at long distance scales. 

As the lightest explicitly non-valence quark, the strange quark
provides an attractive tool to probe the $q\overline{q}$ sea: there
being no valence strange quarks, any strange quark contributions must
be effects of the sea. Thus a quark flavor decomposition of the various
properties of the nucleon becomes of significant interest.  
In
particular, a prominent open question is the 
strange quark contributions to the
distributions of charge and
magnetization.

The use of weak neutral current interactions as key to providing a quark
flavor separation of nucleon currents has been discussed for nearly 
20 years~\cite{Kaplan:1988ku}. The $Z^0$ boson interaction 
with the nucleon is described using form factors which are 
sensitive to a different linear combination of the light quark 
distributions than arise in the more familiar electromagnetic 
form factors. Thus, when combined with electromagnetic form factor
data for the nucleon and the assumption of charge symmetry, 
neutral current measurements allow the 
disentangling of the contributions of the $u,d$ and $s$ 
quarks~\cite{Beck:1989tg,Musolf:1993tb,REVIEWS}.

Recently, experimental techniques 
have developed to the point of enabling measurements of sufficient
precision to access strange-quark effects. The strange quark contributions
to the charge and magnetization of the nucleon are encoded in the 
strange electric and magnetic form factors, $G^s_E$ and $G^s_M$,
analogs of the usual Sachs form factors $G_E$ and $G_M$.

The neutral current interaction can be accessed using parity-violating
electron scattering, in which longitudinally polarized electrons are 
scattered from unpolarized targets. The cross section asymmetry
$A_{\rm PV} = (\sigma_R-\sigma_L)/(\sigma_R+\sigma_L)$ is formed, where 
$\sigma_{R(L)}$ is the cross section for right(left) handed electrons. 
This asymmetry, while typically tiny, of order a few parts per million
(ppm), is caused by the interference of the weak and 
electromagnetic amplitudes, and so it isolates the neutral current
form factors. 

Recently, results of parity-violating electron scattering measurements
on the proton at forward 
angles~\cite{Aniol:1998pn, Maas:2004ta, Maas:2004dh}, 
and on the proton and deuteron at backward 
angles~\cite{Spayde:2003nr} have been
reported.  Each of these individual experiments is sensitive to
different linear combinations of $G^s_E$, $G^s_M$ and the axial
form factor $G^{Zp}_A$.

No individual experiment shows compelling evidence for non-zero strange 
quark effects. However, many available model calculations predicting 
significant strange form factors are allowed by the data. 
It is desirable to carry out complementary measurements 
that could help disentangle the contributions from the various form 
factors. In this paper, we report on experiment E00114, the first 
measurement of $A_{\rm PV}$ for a $^4$He target,
which is sensitive to just one of 
the form factors: $G^s_E$, {\em i.e.} 
the strange quark charge distribution in the 
nucleon~\cite{Beck:1989tg,Musolf:1994gr}.

Elastic electron scattering from $^{4}$He is an 
isoscalar $0^{+}\!\rightarrow\!
0^{+}$ transition and therefore allows no contributions from magnetic or
axial-vector currents. 
The parity-violating asymmetry at tree-level is given 
by~\cite{Musolf:1993tb}
\begin{eqnarray}
A^{\rm He}_{\rm PV} &=&\frac{G_{F}Q^2}
{4\pi \alpha\sqrt{2}} \left( 4\sin ^{2}\theta _{W}+\frac{G_{E}^{s}}{
G_{E}^{\gamma T=0}}\right) \;\;\; ,
\end{eqnarray}
where $G_{E}^{\gamma T=0} = (G_E^{\gamma p}+G_E^{\gamma n})/{2}$ 
is the isospin-zero electric form factor, which is adequately known
from other experiments, 
and $G_{E}^{s}$ is the electric strange quark form factor of the
nucleon. $G_F$ is the Fermi constant, $\alpha$ the fine structure constant,
$\theta _{W}$ the weak mixing angle, and $Q^2$ the square of the
4-momentum transfer. 
The same one-body transition densities appear
in the matrix elements of the weak and electromagnetic operators. When the
ratio comprising an asymmetry is formed, these transition densities cancel \
out, as long as two-body (meson-exchange) currents are negligible. Thus
the nuclear many-body physics divides out  and only the single
nucleon form factors $G_{E}^{s}$ and $G_{E}^{\gamma T=0}$ 
remain~\cite{Musolf:1993tb}.
Nuclear model-dependence in $A^{\rm He}_{\rm PV}$ due to isospin-mixing 
\cite{Ramavataram:1994am} and D-state admixtures~\cite{Musolf:1993mv} in
the $^4$He ground state is negligible, as are 
meson exchange current contributions at the low $Q^2$~\cite{Musolf:1994gr} of
the present experiment.

The experiment was performed in Hall A at the Thomas Jefferson
National Accelerator Facility. A \mbox{$\sim35~\mu$A} continuous-wave
beam of longitudinally polarized 3.03~GeV electrons is incident on a
20~cm long cryogenic high-pressure $^4$He gas target.  Scattered
electrons with $\theta_{lab} \sim  6^{\circ}$ are
focused by two identical spectrometers onto total-absorption
detectors.  Only the electrons are detected from each scattering
event; the second spectrometer merely doubles the accepted solid
angle and provides some cancellation of systematic effects.  
The experimental approach 
largely parallels that of our
previous measurement of parity-violating electron scattering on
hydrogen at higher $Q^2$, which is detailed in~\cite{Aniol:2004hp}.

The spectrometer systems combine the Hall A High Resolution
Spectrometers (HRS)~\cite{Alcorn:2004sb} with new superconducting septum
magnets~\cite{SEPTA}. The septa are necessary to deflect the
$6^{\circ}$ trajectories into the minimum accepted central angle for
the HRS of $12.5^{\circ}$.  
Elastic trajectories are focused onto the detectors, spatially well-separated
from inelastic trajectories by the 12 m dispersion of the HRS.
The elastic trajectories are intercepted by detectors composed of
alternating layers of brass and quartz, oriented such that
Cerenkov light generated by the electromagnetic shower
is transported by the quartz to one end of the detector, to be
collected by a photomultiplier tube (PMT).

The polarized electron beam originates from a GaAs photocathode
excited by circularly polarized laser light.  High polarization 
($\sim85\%$) of the extracted electron beam is achieved using 
an engineered superlattice of doped GaAs semiconductor 
layers~\cite{SUPERLATTICE}.  The helicity of the beam is
set every 33.3 ms locked to the 60 Hz AC power line frequency; 
we refer to each of these periods of constant helicity as a
``window''.  The helicity sequence consists of pairs of
consecutive windows with opposite helicity; the helicity of the first
window in each pair is determined by a pseudo-random number generator.
The response of beam monitors and the electron detector PMTs is
integrated over 
each window 
and then digitized by custom-built analog-to-digital
converters.

The beam current is measured in the experimental hall
with two independent RF cavities and the position of the beam 
is measured at multiple locations using RF strip-line monitors. Typical
intensity jitter at the 33 ms time scale is 600 ppm and 
position jitter on target is 20 $\mu$m.
Careful attention was given in design, component selection, and
configuration of the laser optics in the polarized source
to avoid introducing a false asymmetry due to helicity-correlated
changes in the electron beam properties.
Averaged over the data-taking, the helicity-correlated
asymmetries in the electron beam 
were maintained below 0.075 ppm in intensity, 
0.005 ppm in energy, 7 nm in position and 4
nrad in angle. 

The data sample consists of roughly 3 million pairs of windows, 
corresponding to a 60 hour period. 
For each window, a distributed data acquisition system 
(DAQ) collects data from the polarized source electronics and the 
integrated response of the beam monitors and the detectors. 
Information on the helicity of the beam, delayed
by 8 windows, is included in the data stream. To protect 
against false asymmetry from electronic pick-up, no signal 
carries the helicity information away from the source region 
without this 8 window delay.

The only cuts applied to the data are to remove periods of 
either low beam current or rapidly changing current, or 
when a spectrometer magnet is off. No helicity-dependent cuts are applied.

The helicity-dependent asymmetry in the integrated detector response,
normalized to the average beam intensity during each window, is
computed for each pair of windows and then corrected for fluctuations in
the beam trajectory to form the raw asymmetry $A_{\rm raw}$.  To first
order, five correlated parameters describe the average trajectory of
the beam during a window: energy, and horizontal and vertical positions
and angles.

Two independent methods are used to calibrate
the detector sensitivity to each beam parameter, and thus
remove the beam-induced random and systematic effects from
the raw detector-response asymmetry.  The first uses a 
calibration subset of helicity windows, where each beam
parameter is modulated periodically around its average
value by an amount large compared to nominal beam fluctuations.
The other method applies
linear regression to the window pairs used in the physics
analysis.  

These techniques yield results which differed by
a negligible amount compared to the final statistical error.
Final results are obtained using the
modulation technique.
The cumulative correction for $A_{\rm raw}$ due to beam trajectory and 
energy asymmetry is $-0.026$ ppm. 

The small beam intensity asymmetry of 0.075 ppm would induce a false
asymmetry proportional to any alinearity in the detectors and beam 
current monitors.  
Using dedicated runs, the alinearity of the detector system is 
determined to be less than 1\% and the relative alinearity between 
the beam monitors and the detectors is found to be less 
than 0.2\%.  

\begin{figure}
\begin{center}
\includegraphics[width=3.5in]{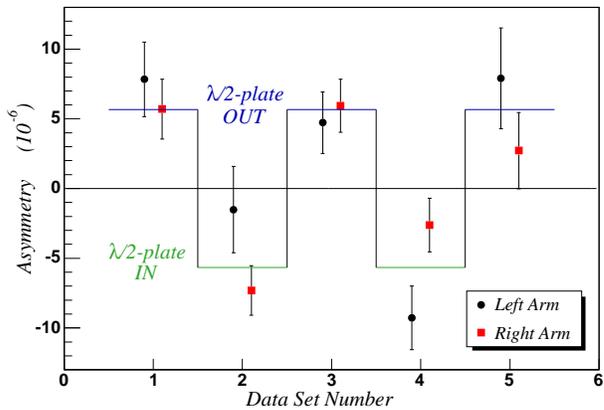}
\caption{Raw detector asymmetry $A_{\rm raw}$ for both spectrometers, broken down 
by data set. The step pattern represents the insertion/removal of a 
$\lambda/2$ plate at the beam source, which should flip the 
sign of the measured asymmetry.}
\label{figure:slug} 
\end{center}
\end{figure}

A half-wave ($\lambda$/2) plate is periodically inserted into the 
source laser optical path. This passively reverses the
sign of the electron beam polarization, and hence the sign of
$A_{\rm raw}$, while leaving many possible systematic
effects unchanged. Roughly equal statistics were accumulated 
with and without the $\lambda$/2 plate, thereby suppressing 
many systematic effects.   Figure~\ref{figure:slug} shows 
$A_{\rm raw}$ for all the data, grouped by periods of constant 
$\lambda$/2-plate state.

The physics asymmetry $A^{\rm He}_{\rm PV}$ is formed from $A_{\rm raw}$ 
by correcting for beam polarization, backgrounds, and finite acceptance:
\begin{equation}
A^{\rm He}_{\rm PV} = \frac{K}{P_b}\frac{A_{\rm raw} - P_b\sum_{i} 
A_{i}f_{i}}{1-\sum_{i} f_{i}} \;\;\; ,
\end{equation}
where $P_b$ is the beam polarization, $f_{i}$ are background fractions and
$A_{i}$ the associated background asymmetries,
and $K$ accounts for the range
of kinematic acceptance.

The beam polarization is measured in the experimental hall using a
Compton polarimeter~\cite{ESCOFFIER} which provides a continuous,
non-invasive measurement simultaneous with data collection. Averaged
over the run, the polarization is determined to be $P_{b} = (86.9 \pm
1.7)\%$.  This result is consistent within error with results from
dedicated runs to measure polarization using a M{\o}ller
polarimeter~\cite{Alcorn:2004sb}.

Tracking chambers, part of the standard HRS detector
package~\cite{Alcorn:2004sb}, are used to track individual events at the
focal plane during dedicated, low-current runs in order  
to determine the average kinematics and to 
study backgrounds to the integrating measurement.

\begin{figure}
\begin{center}
\includegraphics[width=3.0in]{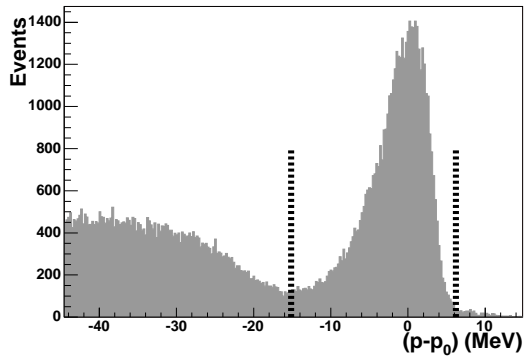}
\caption{Measured momentum difference from the central momentum ($p_0$)
of one spectrometer, at the focal plane (filled 
histogram).
Quasielastic scattering from $^4$He dominates the data
at low momenta. The vertical lines indicate the edges of the
Cerenkov detector acceptance.}

\label{figure:deltap} 
\end{center}
\end{figure}

The total background is found to comprise $<3\%$ of the 
detector signal, of which inelastic scattering from $\He$ is the 
largest component.
The inelastic fraction is determined to be  $(1.6 \pm 0.8)\%$ of the total
detected flux by extrapolating the rise of the inelastic signal above the
elastic radiative tail into the low-momentum edge of the 
detector --- see Figure~\ref{figure:deltap}. This contribution is 
dominated by quasi-elastic (QE) scattering.

The rescattering of electrons from various spectrometer apertures
is another source of background. This is studied 
by varying the central spectrometer momentum in dedicated runs,
and is determined to be 
$(0.6 \pm 0.6)\%$, dominated by QE scattering.  
Contributions from exposed iron
pole-tips in the dipole are negligible.

Background due to the aluminum windows of the cryogenic target is
measured using an aluminum target with thickness matching the
radiation length of the full target cell. The background is $(0.7 \pm
0.1)\%$ of the total detected rate; studies at low current to track
individual counting events show that this contribution is dominated
by QE scattering.

The corrections to the measured asymmetry for the QE scattering 
backgrounds discussed above
are determined using the predicted~\cite{Musolf:1992xm}
asymmetry of $-1.6$ ppm for both aluminum and $^4$He, with a conservative 
50\% error assumed. The background corrections to $A_{\rm raw}$ 
are small and are listed in Table~\ref{table:Acorrections}.

The average $Q^{2}$ is determined 
to be $\langle Q^{2} \rangle =  (0.091 \pm 0.001) \; {\rm GeV}^2$ by dedicated
low-current runs. Determination of $Q^2$ at this level requires  
precision measurement of the absolute scattering angle. 
Due to nuclear recoil, the scattering angle can be determined from the 
momentum difference between electrons elastically scattered 
from hydrogen and from a heavy nucleus. 
A  water-cell target provided a target containing 
hydrogen and the heavier oxygen nuclei.  
The scattering angle into pinholes 
of a sieve collimator at the entrance of each spectrometer is 
measured using this method to a 
precision of 0.3\%.

\begin{table}
\begin{tabular}{lrcl}
\multicolumn{4}{l}{Correction (ppm)}\\ \hline \hline
Target windows & $0.058$  & $\pm$ & $0.012$ \\
QE $\He$ &       $0.129$  & $\pm$ & $0.070$ \\
Rescattering &      $0.049$  & $\pm$ & $0.050$ \\
Beam Asyms. &    $-0.026$ & $\pm$ & $0.102$  \\
Alinearity &     $0.000$   & $\pm$ & $0.077$  \\
\hline \hline \\
\multicolumn{4}{l}{Normalization Factors}\\ \hline \hline
Polarization $P_b$ & $0.869$ & $\pm$ & $0.017$ \\
Acceptance $K$     & $1.000$ & $\pm$ & $0.001$ \\
$Q^{2}$ Scale      & $1.000$  & $\pm$ & $0.010$ \\ \hline \hline
\end{tabular}
\caption{Corrections to $A_{\rm raw}$ and systematic uncertainties.
\label{table:Acorrections}}
\end{table}

Results from the two spectrometers agree within the statistical errors and
are averaged together.
After all corrections, the asymmetry is found to be 
$A^{\rm He}_{\rm PV} = 6.72 \pm 0.84 \pm 0.21$~ppm, 
where the first error is statistical and the second systematic. Individual
contributions to the systematic error are detailed in 
Table~\ref{table:Acorrections}. 

The theoretical value for the asymmetry from Eq.~1, under the assumption
that $G^s_E =0$, including the (small) vector electroweak radiative 
corrections~\cite{Musolf:1993tb}, 
is $A^{\rm He}_{\rm PV}|_{G^s_E=0} = 7.483\;  {\rm ppm}$. 
The effect of purely electromagnetic radiative corrections is negligible
due to the spin independence of soft photon emission and the small
momentum acceptance of the detectors.  
For the electromagnetic form factor $G_{E}^{\gamma T=0}$
we have used a recent phenomenological fit to the world data at 
low $Q^2$~\cite{Friedrich:2003iz}, with a total uncertainty of 2.6\%.
Comparing $A^{\rm He}_{\rm PV}|_{G^s_E=0}$ to our measured 
$A^{\rm He}_{\rm PV}$ we extract
the value of the strange electric form factor
$G^s_E = -0.038 \pm 0.042 \pm  0.010$, which is consistent with zero. 
The first uncertainty is statistical and the second is systematic,
including those due to radiative corrections and electromagnetic
form factors. 

There have been numerous attempts to calculate strange form factors
using a host of models and theoretical approaches (see~\cite{REVIEWS}
and references therein). Available calculations do not even agree on
the sign of $G^s_E$, and predicted values range from $-0.08$ to $+0.08$. 
The present result disfavors models with large positive values, e.g.
\cite{Weigel:1995jc,Silva:2001st}. 

We have also made a measurement of $A_{\rm PV}$ from the proton 
at a very similar $Q^2$, which is reported in an accompanying 
paper~\cite{HAPPEXII}. The combination of the two measurements, as well
as
with previous measurements at the same 
$Q^2$~\cite{Maas:2004dh,Spayde:2003nr} allows 
access to both $G^s_E$ and $G^s_M$ separately~\cite{HAPPEXII}. 
A new run of this experiment is scheduled for Summer 2005, which 
is expected to improve the statistical precision by a factor of 
3, along with a modest reduction in the systematic error.  

In summary, we have made the first measurement of the parity-violating 
asymmetry in elastic electron scattering from $^4$He, which is uniquely
sensitive to the strange electric form factor $G^s_E$. The result
obtained is consistent with zero, and constrains models of the strangeness
in the nucleon.

We thank the superb staff of Jefferson Lab for their contributions 
to the success of this work.
This work was supported 
by DOE contract DE-AC05-84ER40150, Modification No. M175
under which the Southeastern Universities 
Research Association
(SURA) operates the Thomas Jefferson 
National Accelerator Facility, 
and by the Department of 
Energy, the National Science Foundation, 
the Istituto Nazionale di Fisica Nucleare (Italy), 
the Commissariat \`a l'\'Energie Atomique (France), 
and the Centre National
de Recherche Scientifique (France).

\end{document}